\documentclass[CEJP,PDF,DVI]{cej}
\usepackage{layout}
\usepackage{amsmath}
\usepackage{textcomp}
\usepackage{hyperref}
\usepackage{dcolumn}

\title{Accurate calculation of the eigenvalues of a new simple class of superpotentials
in SUSY quantum mechanics}

\articletype{Research Article}

\author{Francisco M. Fern\'{a}ndez\inst{1} \email{fernande@quimica.unlp.edu.ar}}

\institute{\inst{1} INIFTA (UNLP, CCT La Plata-CONICET),
Divisi\'on Qu\'imica Te\'orica, Blvd. 113 S/N,  Sucursal 4,
Casilla de Correo 16, 1900 La Plata, Argentina}

\abstract{ We obtain accurate eigenvalues for two  recently derived SUSY partner
Hamiltonians. We improve the Rayleigh-Ritz
variational method proposed by the authors and show how to apply
the Riccati-Pad\'{e} method to those particular partner
potentials.}



\begin{document}

\maketitle

\section{Introduction}

\label{sec:intro}

In a recent paper Marques et al\cite{MNS12} derived two new simple
supersymmetric partner potentials and showed that the ground-state
eigenfunction of one of the Hamiltonian operators can be obtained exactly.
In order to calculate the remaining states they resorted to the
Rayleigh-Ritz variational method. However, the convergence rate of their
approach appears to be insufficient to reveal the relation between
the eigenvalues of the partner Hamiltonians clearly.

The authors also estimated the ground-state energy of the quartic anharmonic
oscillator by means of perturbation theory starting from the known solution
for one of the partner Hamiltonian operators.

The purpose of this paper is twofold. First we improve the
variational method proposed by Marques et al\cite{MNS12} and,
second, we show that the Riccati-Pad\'{e} method
(RPM)\cite{FMT89a,FMT89b,F08} provides highly accurate results for
the eigenvalues of those partner Hamiltonian operators. In
addition, we show that exactly the same Rayleigh-Ritz
variational method applied to the SUSY partner Hamiltonians is
also suitable for the calculation of the eigenvalues of the
quartic anharmonic oscillator.

In Sec.~\ref{sec:SUSY} we apply the Rayleigh-Ritz variational
method and the RPM to the pair of SUSY Hamiltonians proposed by
Marques et al\cite{MNS12} and also to the quartic oscillator. In
Sec.~\ref {sec:conclusions} we comment on those approaches and
draw conclusions.

\section{SUSY partners}

\label{sec:SUSY}

The partner Hamiltonian operators are given by\cite{MNS12}
\begin{equation}
H_{\pm }=-\frac{d^{2}}{dx^{2}}+x^{4}\pm 2|x|  \label{eq:Hamiltonians}
\end{equation}
where we have chosen $g=1$ without loss of generality.

In order to solve the Schr\"{o}dinger equation
\begin{equation}
H_{\pm }\psi _{n}^{\pm }=E_{n}^{\pm }\psi_{n}^{\pm}
\label{eq:Schro}
\end{equation}
approximately, Marques et al\cite{MNS12} chose the nonorthogonal
basis set
\begin{equation}
\varphi _{j}(x)=x^{j-1}e^{-|x|^{3}/3},\,j=1,2,\ldots  \label{eq:Basis_MNS}
\end{equation}
and derived a single secular equation for both the even and odd
states. Since states of different parity do not mix the dimension
of their secular equation is twice the size of what would be
actually necessary for a calculation of the same accuracy.

In this paper we propose to work on the half line $x\in [0,\infty
)$ taking into account the appropriate boundary conditions at
the origin: $\psi (0)\neq 0$, $\psi ^{\prime }(0)=0$ for the even
states and $\psi (0)=0$, $\psi ^{\prime }(0)\neq 0$ for the odd
ones. The appropriate basis sets are
\begin{eqnarray}
\varphi _{j}^{e} &=&x^{j}e^{-x^{3}/3},\,j=0,2,3,\ldots  \nonumber \\
\varphi _{j}^{o} &=&x^{j}e^{-x^{3}/3},\,j=1,2,3,\ldots  \label{eq:Basis_mine}
\end{eqnarray}
for the former and latter case, respectively.

Tables \ref{tab:RR1} and \ref{tab:RR2} show Rayleigh-Ritz results
for the first eigenvalues of $H_{-}$ and $H_{+}$, respectively, in
terms of the size $N$ of the basis sets. We appreciate that the
variational results decrease with the number of basis functions as
expected. However, we are not aware of any proof that either of
the basis sets (\ref{eq:Basis_MNS}) or (\ref {eq:Basis_mine}) is
complete. We see that present results are smaller, and therefore
more accurate, than those of Marques et al\cite{MNS12}. In
addition, the results in their tables 1 and 2 exhibit
unchanged entries when the size of the basis set increases from
$m$ to $m+1$ which reflects the fact that the added basis function
improves the even (odd) state but has no effect on the odd (even)
one. As a result their matrices are twice the size of what is
necessary for a calculation of the same accuracy.

The results of Tables \ref{tab:RR1} and \ref{tab:RR2}, as well as
those of Marques et al\cite{MNS12}, suggest that
$E_{n}^{-}=E_{n-1}^{+}$, $n=1,2,\ldots $, as expected for a SUSY
pair of partner Hamiltonian operators. However, those variational
results are not sufficiently accurate and, as argued above, we are
not aware of any proof of convergence. In order to obtain more
accurate results we resort to the RPM\cite{FMT89a,FMT89b,F08}. To
this end we consider the modified logarithmic derivative of the
wavefunction $\psi $
\begin{equation}
f(x)=\frac{s}{x}-\frac{\psi ^{\prime }(x)}{\psi (x)}  \label{eq:log_der}
\end{equation}
where $s=0$ or $s=1$ for even or odd states, respectively. It satisfies the
Riccati equation
\begin{equation}
f^{\prime }(x)=f(x)^{2}-\frac{2s}{x}f(x)+E-x^{4}\mp 2x,\,x>0
\label{eq:Riccati}
\end{equation}
and can be expanded in a Taylor series about the origin
\begin{equation}
f(x)=\sum_{j=0}^{\infty }f_{j}x^{j}  \label{eq:f_expansion}
\end{equation}
where the coefficients $f_{j}$ can be easily obtained from the
Riccati equation (\ref{eq:Riccati}). We have $f_{0}=-\psi ^{\prime
}(0)/\psi (0)=0$ for the even states and choose $f_{0}=0$ for the
odd ones in order to remove the unbalanced term $2f_{0}/x$ that
appears when we substitute the expansion (\ref {eq:f_expansion})
into the Riccati equation (\ref{eq:Riccati}). Note that this
application of the RPM to even potentials is different from the
one in earlier papers\cite{FMT89a,FMT89b,F08} because in the
present case we restrict the calculation to the half line. The
reason for such a
modification is that the present even potentials cannot be expanded in an $x^{2}$%
-power series.

With the coefficients of the expansion (\ref{eq:f_expansion}) we construct
the Hankel determinants $H_{D}^{d}=\left| f_{j+j+d-1}\right| _{i,j=1}^{D}$
that are polynomial functions of the energy and obtain the approximate
eigenvalues from the roots of $H_{D}^{d}(E)=0$\cite{FMT89a,FMT89b,F08}. More
precisely, we expect that sequences of roots $E_{n}^{[D,d]}$ converge
towards the eigenvalues $E_{n}$, $n=0,1,\ldots $, of the corresponding
Hamiltonian operator as $D$ increases. Since the SUSY partner potentials on
the half line $V_{\pm }(x)=x^{4}\pm 2x$ satisfy $V_{+}(-x)=V_{-}(x)$ then
the RPM yields the eigenvalues of both partner Hamiltonians simultaneously%
\cite{F96}. More precisely, the sequences of roots for the even
and odd wavefunctions will converge towards
$E_{0}^{-}<E_{0}^{+}<E_{2}^{-}<E_{2}^{+}<\ldots $ and
$E_{1}^{-}<E_{1}^{+}<E_{3}^{-}<E_{3}^{+}<\ldots $, respectively.

Table \ref{tab:RPM} shows the lowest eigenvalues for the SUSY
partner Hamiltonians estimated from Hankel sequences with $D\leq
30$, $d=0$ and $d=1$. The RPM yields the ground-state eigenvalue
of $H_{-}$ exactly ($E_{0}^{-}=0 $) and there is no doubt from
present accurate eigenvalues that $E_{n}^{-}=E_{n-1}^{+}$.

Marques et al\cite{MNS12} estimated the ground-state eigenvalues
for the Schr\"{o}dinger equation with the potentials $V(x)=x^{4}$
and $V_{+}(x)$ from the exact ground-state solution $\psi
_{0}^{-}$, $E_{0}^{-}$ for $H_{-}$. They resorted to logarithmic
perturbation theory but their results were not accurate. Here, we
point out that they could have obtained more accurate results by
means of the variational method discussed above. In Table
\ref{tab:RRx4} we show the results coming from the more efficient
Rayleigh-Ritz variational method proposed in this paper and we
also add the accurate eigenvalues produced by the RPM for
comparison.

\section{Conclusions}

\label{sec:conclusions}

It is well known that the Rayleigh-Ritz variational method is more
efficient if one treats every symmetry species separately. In the
present case we have carried out the separation into even and odd
functions on the half line by means of appropriate boundary
conditions at origin. Thus, the size of present secular equations,
and consequently the computational effort for a given accuracy, is
one half that of Marques et al\cite{MNS12}.

Working on the half line and taking into account the boundary
conditions at origin also enables one to obtain highly accurate
eigenvalues by means of the RPM. In the present case one has to be
cautious because the same Hankel determinant exhibits roots close
to the eigenvalues of both partner Hamiltonian operators. However,
it is not difficult to identify the sequence of roots that
converges to a chosen eigenvalue. The present RPM results confirm that
the two Hamiltonian operators proposed by Marques et al\cite
{MNS12} are already SUSY partners.

The RPM was applied to sequences of SUSY partner potentials in the past\cite
{FMDT89}. However, the main interest in that paper was to obtain
excited-state eigenvalues with Hankel determinants of relatively small
dimension. Besides, those potentials can be expanded in $x^{2}$-power series
and can therefore be treated in the usual way.

\begin{table}[H]
\caption{Eigenvalues $E_n^-$ from the Rayleigh-Ritz method}
\label{tab:RR1}
\begin{center}
\par
\begin{tabular}{llllll}
\hline
$N$ & \multicolumn{1}{c}{$n=0$} & \multicolumn{1}{c}{$n=1$} &
\multicolumn{1}{c}{$n=2$} & \multicolumn{1}{c}{$n=3$} & \multicolumn{1}{c}{$%
n=4$} \\ \hline
2 & 0 & 1.970246841 & 5.765408776 & 9.488542554 &  \\
3 & 0 & 1.970246137 & 5.511879703 & 9.488418664 & 14.13007837 \\
4 & 0 & 1.969640134 & 5.507908922 & 9.406978612 & 13.89266195 \\
5 & 0 & 1.969515816 & 5.507440820 & 9.394523525 & 13.87435060 \\
6 & 0 & 1.969507628 & 5.507202146 & 9.394324659 & 13.85957259 \\
7 & 0 & 1.969507538 & 5.507178381 & 9.394287127 & 13.85838650 \\ \hline\hline
&  & \multicolumn{1}{c}{$n=5$} & \multicolumn{1}{c}{$n=6$} &
\multicolumn{1}{c}{$n=7$} & \multicolumn{1}{c}{$n=8$} \\ \hline
2 &  &  &  &  &  \\
3 &  & 19.48962926 &  &  &  \\
4 &  & 19.02107688 & 24.09989393 & 33.11107172 &  \\
5 &  & 18.66009505 & 24.07406802 & 29.81927599 & 35.91298387 \\
6 &  & 18.64895855 & 23.84777162 & 29.26028194 & 35.65656448 \\
7 &  & 18.64704858 & 23.80790714 & 29.25748454 & 34.98577701 \\ \hline
\end{tabular}
\par
\end{center}
\end{table}

\begin{table}[H]
\caption{Eigenvalues $E_n^+$ from the Rayleigh-Ritz method}
\label{tab:RR2}
\begin{center}
\par
\begin{tabular}{lllll}
\hline
$N$ & \multicolumn{1}{c}{$n=0$} & \multicolumn{1}{c}{$n=1$} &
\multicolumn{1}{c}{$n=2$} & \multicolumn{1}{c}{$n=3$} \\ \hline
2 & 1.970246841 & 5.529040685 & 9.488542554 & 14.35721210 \\
3 & 1.970246137 & 5.514418950 & 9.488418664 & 14.00103534 \\
4 & 1.969640134 & 5.510107538 & 9.406978612 & 13.92544875 \\
5 & 1.969515816 & 5.507493533 & 9.394523525 & 13.86389218 \\
6 & 1.969507628 & 5.507185747 & 9.394324659 & 13.85851252 \\
7 & 1.969507538 & 5.507178915 & 9.394287127 & 13.85851126 \\ \hline\hline
& \multicolumn{1}{c}{$n=4$} & \multicolumn{1}{c}{$n=5$} & \multicolumn{1}{c}{%
$n=6$} & \multicolumn{1}{c}{$n=7$} \\ \hline
2 &  &  &  &  \\
3 & 19.48962926 & 24.52838046 &  &  \\
4 & 19.02107688 & 24.52809061 & 33.11107172 & 37.63335229 \\
5 & 18.66009505 & 23.91353748 & 29.81927599 & 36.33167441 \\
6 & 18.64895855 & 23.81074205 & 29.26028194 & 35.02577244 \\
7 & 18.64704858 & 23.81074194 & 29.25748454 & 34.98424967 \\ \hline
\end{tabular}
\par
\end{center}
\end{table}

\begin{table}[H]
\caption{Eigenvalues for the partner SUSY Hamiltonian operators from the
Riccati-Pad\'e method}
\label{tab:RPM}
\begin{center}
\par
\begin{tabular}{D{.}{.}{20}D{.}{.}{20}}
\hline

\multicolumn{1}{c}{Even ($s=0$)} &
\multicolumn{1}{c}{Odd ($s=1$)}  \\
\hline

0                       &     1.9695075137502948249  \\
 1.9695075137502948249  &     5.5071777771459699676   \\
 5.5071777771459699676  &     9.3942674378738914658   \\
 9.3942674378738914658  &    13.858371936541300147    \\
13.858371936541300147   &    18.645975633988444799    \\
18.645975633988444799   &    23.807185917985766918    \\
23.8071859179857669     &    29.2325545506214961      \\
29.23255455062149608    &    34.9463357188906106      \\
34.94633571889061       &                             \\

\hline

\end{tabular}
\par
\end{center}
\end{table}

\begin{table}[H]
\caption{Eigenvalues $E_n$ for the quartic oscillator}
\label{tab:RRx4}
\begin{center}
\par
\begin{tabular}{llll}
\hline
$N$ & $n=0$ & $n=1$ & $n=2$ \\ \hline
2 & 1.077335422 & 3.804924324 & 8.102531212 \\
3 & 1.061889825 & 3.802939362 & 7.486487866 \\
4 & 1.060417725 & 3.800572274 & 7.456738018 \\
5 & 1.060362727 & 3.799746874 & 7.456540880 \\
6 & 1.060362727 & 3.799674368 & 7.455826496 \\
7 & 1.060362223 & 3.799673299 & 7.455703721 \\ \hline
RPM & 1.0603620904841828996 & 3.7996730298013941688 & 7.4556979379867383922
\\ \hline\hline
& $n=3$ & $n=4$ & $n=5$ \\ \hline
2 & 11.86759677 &  &  \\
3 & 11.77079816 & 16.91718138 & 21.92232689 \\
4 & 11.67894913 & 16.29307103 & 21.79883488 \\
5 & 11.64634787 & 16.29029658 & 21.28503933 \\
6 & 11.64483615 & 16.26577116 & 21.24138553 \\
7 & 11.64480669 & 16.26191320 & 21.24054440 \\ \hline
RPM & 11.644745511378162021 & 16.261826018850225938 & 21.238372918235940024
\\ \hline
\end{tabular}
\par
\end{center}
\end{table}

\end{document}